# Reconfigurable Spin Logics and High-density Multistate Memory in a Single Spin-orbit Torque Device

Raghvendra Posti[1], Dhanajay Tiwari[2], and Debangsu Roy[1*]

[1]Department of Physics, Indian Institute of Technology Ropar, Rupnagar 140001, India

[2]Silicon Austria Labs GmbH, Sandgasse 34, Graz 8010, Austria

Nonvolatile devices based on the spin-orbit torque (SOT) mechanism are highly suitable for in-memory logic operations. The current objective is to enhance the memory density of memory cells while performing logic operations within the same memory unit. Present study demonstrates that integrating SOT with an out-of-plane magnetic field effectively achieves multiple magnetic states in perpendicularly magnetized heterostructures. This study further explores this approach, experimentally demonstrating reconfigurable logic operations within a single SOT device using W/Pt/Co/AlOx heterostructures. Our results show that multistate tuning by SOT integration with out-of-plane magnetic field enables reconfigurable logic operations, including AND, OR, NOR, NAND, and Always ON, within a single device. Additionally, we found that careful selection of input logic operations allows multiple configurations to achieve the same logic function within a single memory device. To enhance multistate memory density, we proposed and experimentally verified a two-step writing process, achieving the highest reported multistate memory density in SOT-based memory devices. These findings highlight the potential of integrating SOT and magnetic field effects to realize high-density, multifunctional in-memory logic devices.

* Corresponding author: debangsu@iitrpr.ac.in

In contemporary computing, Von Neumann architecture involves data flowing through a memory hierarchy to the processor[1]. This shuttling of data from the processor to memory consumes a considerable amount of energy and incurs latency in data processing. To overcome these drawbacks, an alternative method is in-memory computing, where both storage and logic operations occur within the memory unit itself. To achieve these operations in a single memory cell, the memory cells must have the multifunctionality of memory and logic devices. Due to their non-volatile nature, fast access time, small footprints, and low energy consumption, spin-orbit torque (SOT) based magnetic memory devices are widely studied for these applications[2-4]. To perform in-memory computing, SOT-based memory elements need to have multiple memory states instead of binary states. Multistate memory behavior in these devices helps achieve reconfigurable logic functions within them. Therefore, recent efforts have focused on increasing the memory density of these memory elements and employing these multi-states to demonstrate reconfigurable logic operations[5-15]. However, the multifunctionality of reconfigurable logic devices and multistate memory is often the result of unconventional device and/or material heterostructures. For example, a recent study by Y. Fan et al. demonstrated multiple hysteresis configurations for multistate and reconfigurable logic operations by tuning in-plane and out-of-plane exchange bias[13]. However, the complex material stack and the need to access out-of-plane exchange bias below room temperature make this method challenging to implement. Additionally, various other reports have showcased reconfigurable logic operations in single SOT-based devices through unconventional device designs or complex experimental setups, such as applying current in two orthogonal directions[11,14,16]. Moreover, advanced material engineering techniques, including gradient composition of heavy metals[10] and precise control of exchange bias and/or interlayer exchange coupling[7,11], have been employed to achieve similar results. Nevertheless, most of these methods still limit the multistate magnetic memory density to a small number.

In our previous study**[REF]**, we demonstrated that the integration of SOT with an out-of-plane magnetic field is an effective and universal method to achieve multiple magnetic states in perpendicularly magnetized heterostructures. This approach could potentially enable the realization of both high-density multi-states and reconfigurable logic operations within a single device. In this study, we experimentally demonstrated reconfigurable logic operations in a single SOT device by integrating SOT and magnetic field. Our experiments with W/Pt/Co/AlOx show that in a heavy metal (HM)/ferromagnet (FM) heterostructure with perpendicular magnetic anisotropy (PMA), multistate tuning can enable reconfigurable logic operations such as AND, OR, NOR, NAND, and Always ON within a single SOT device. Additionally, we show that by carefully selecting input logic operations, it is possible to achieve a single logic function in multiple ways within a single memory device.In addition to in-memory logic operations, achieving high-density multi-states within a single memory cell is crucial for significantly enhancing memory performance. To further increase multistate memory density, a two-step writing process was proposed and successfully verified experimentally.

By employing this two-step writing approach, we achieved the highest multistate memory density reported so far in SOT-based memory devices.

A perpendicularly magnetized multilayer stack of Ta(3)/W(0.85)/Pt(3)/Co(0.6)/AlOx(1.3) was sputter-deposited on a thermally oxidized Si/SiO2 substrate, with the numbers in parentheses indicating layer thickness in nanometers. The sputtering process was conducted at room temperature under a base pressure of $1\times10^{-7}$ mbar, with an Ar gas environment of $2\times10^{-3}$ mbar. The multilayers were then patterned into six-terminal Hall bar devices with a current channel width of approximately 12 μm using laser lithography. The fabricated devices were wire-bonded to external circuitry for room temperature electrical transport measurements. Both AC (SOT characterization) and DC (magnetization switching) experiments were performed using equipment such as a Keithley 6221 current source, a Keithley 2182A nanovoltmeter, and an EG&G 7265 lock-in amplifier.

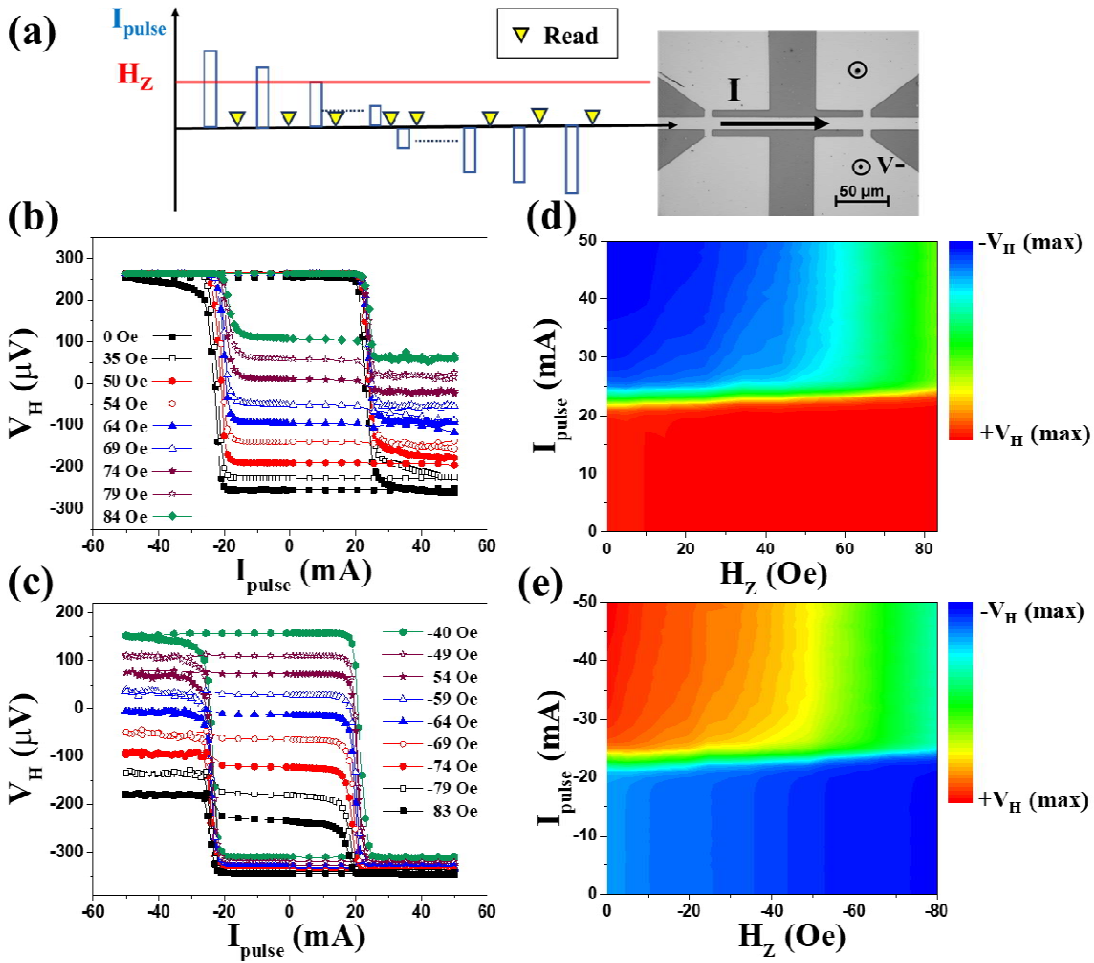

**Fig. 1(a)** A six-terminal Hall bar device with writing current scheme for SOT and $H_Z$ controlled magnetization switching. Writing current includes the current pulse sweep between 45 mA (5 ms width). This writing operation was followed by Hall measurement-based reading mechanism

(voltage measurements at transverse Hall branches) at 1 mA DC. $H_Z$ controlled SOT switching hysteresis for **(b)** positive and **(c)** negative magnitudes of $H_Z$. Color diagram showing Hall voltage reading (magnetization orientation) as a function of current pulses magnitudes as y-axis and x-axis as **(d)** positive $H_z$ values **(e)** negative $H_Z$ values.

When a charge current flows through the HM layer of magnetic heterostructures, it generates a spin current due to the bulk spin-Hall effect (SHE)[17-19] and/or the interfacial Rashba–Edelstein effect[20-22]. This spin current, in turn, induces a torque in the adjacent FM layer. The predominant anti-damping-like component of the SOT is considered responsible for driving the magnetization reversal[23,24] (please refer supplementary information for the detailed characterization of SOT). In our previous study, we theoretically predicted and experimentally demonstrated that in PMA-based heterostructures, the critical SOT-induced field value can be altered by applying an out-of-plane magnetic field during SOT-induced magnetization switching. This alteration in the critical AD-SOT effective field leads to domain state formation, enabling the integration of SOT with an external field along the out-of-plane direction ($H_z$), resulting in the formation of multiple states.

The square hysteresis of the anomalous Hall effect (AHE) signal vs. $H_Z$ verifies the PMA in our deposited stack (see supplementary Fig. S1). Using an external out-of-plane magnetic field to achieve multistate behavior in SOT-driven magnetization switching within PMA-based heterostructures is a universal experimental method [REF]. Therefore, we anticipate our W/Pt/Co/AlOxheterostructure to exhibit similar behavior. To confirm this, we applied 5 ms current pulses ($I_{pulse}$) of varying magnitudes along the current channel of the Hall bar device, subsequently, the magnetization orientation is read by conventional AHE ($V_H$) measurements by utilizing a small reading current of 1 mA. The schematic of the SOT-induced magnetization switching in the presence of a constant $H_Z$ is shown in Fig. 1(a). Note that a symmetry-breaking field ($H_X$) of 1300 Oe was applied in all the experiments performed in this work.During an individual current-induced switching measurement through SOT, a fixed $H_Z$ is maintained. Figures 1(b) and 1(c)show the results of SOT-induced magnetization reversal in the presence of different $H_Z$ magnitudes with positive and negative polarities, respectively.

Here, the magnetization state decreases depending on the magnitude and polarity of the applied $H_Z$ field. A positive (negative) magnitude of $H_Z$favors the $+m_z$ ($-m_z$) state and reduces the critical AD-SOTfields leading to an incomplete hysteresis of the $-m_z$ ($+m_z$) state.This reduction in the magnetic state is observed as a decrease in the AHE signal, which implies the formation of domain states in the device. Along with the multistate nature introduced by $H_Z$and SOT, current-induced hysteresis loops also show a linear shift along the current axis corroborating the theoretical macrospin analysis reported in our previous work[ref].Furthermore, the shift can be explained by the 1D-domain model[25,26], suggesting that the effective field of AD-SOT acts perpendicular to the heterostructureplane and scales linearly with the applied current ($H_{AD} \propto I$). Therefore, an additional $H_Z$ should act as an offset along the current axis in the current-induced hysteresis loop. Both of these features—state reduction and hysteresis loop shift—are notable in the color maps shown in Figures 1(d) and 1(e).Here, the Hall voltage (represented by the color gradient) is plotted as a function of applied current pulses and $H_Z$ values. A linear transition region separates the fixed state (region below $|I_{pulse}| < 20$ mA) from the varying state. Notably, the application of $+H_Z$ fixes the $+m_z$ state (indicated by the red color in Fig. 1(d)), while $-H_Z$ fixes the $-m_z$ state (indicated by the blue color in Fig. 1(e)).

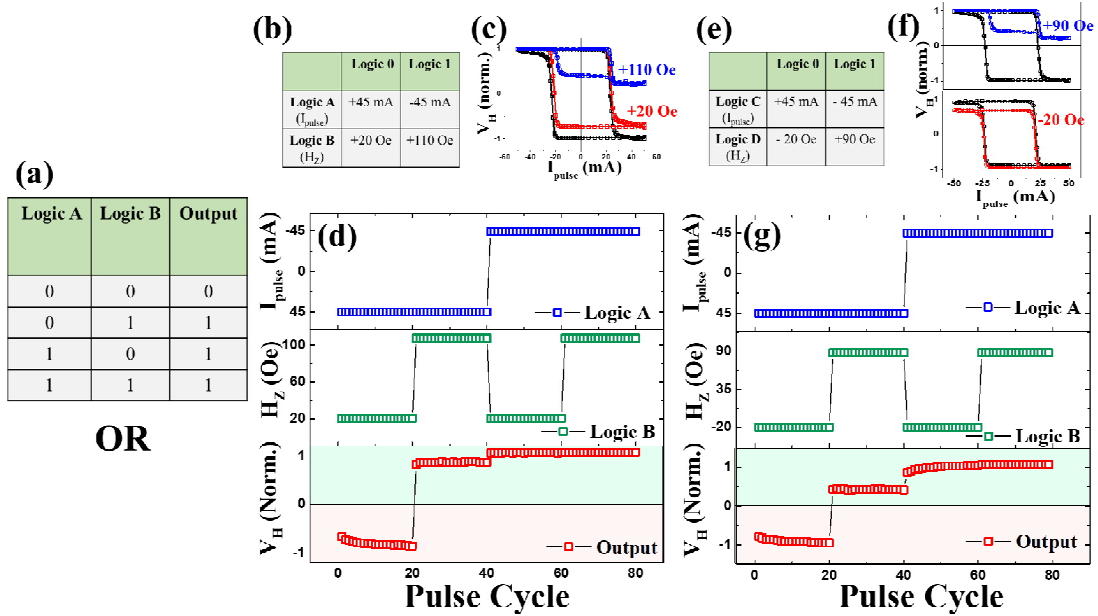

**Fig. 2(a)**Typical OR gate truth table. **(b)**Logic 0 and 1 for input logics A and B**(c)**Current induced hysteresis at field values mentioned in logic B**(d)**Operational sequence of OR gate with input condition mentioned in (b). **(e)**Logic 0 and 1 for input logicsC and D**(f)** Current induced hysteresis at different field values mentioned in logic D **(g)** operational sequence of OR gate with input condition mentioned in (e).

Next, we utilized the multistate behavior of our PMA stack, integrated with an $H_Z$ field, to perform reconfigurable spin logic operations. By carefully setting the logic inputs and initialization conditions, we can achieve a variety of logic operations. Additionally, by selecting specific input configurations, a single logic operation can be validated in multiple ways. In our setup, logic inputs A and Care current-dependent, while logic inputs B and Dare $H_Z$ field-dependent. The output logic is determined by the polarity of the Hall signal: a positive Hall voltage corresponds to a logic output of 1, and a negative Hall voltage corresponds to a logic output of 0. Figures 2 and 3 illustrate the logic operations of OR and AND gates, respectively, under two different sets of input conditions for each case. For all operations, the $H_Z$-controlled multistate behavior in SOT-induced switching is achieved by adjusting the $H_Z$ values, as depicted in Fig. 1. The truth table for an OR gate is shown in Fig. 2(a), with logic inputs A and B detailed in Fig. 2(b). Here, logic A ($I_{pulse}$ current logic) is 0 (1) at the current value of +45 mA (-45mA) and logic B ($H_Z$ field logic) is 0 (1) corresponding to the $H_Z$ field value of +20 Oe (+110 Oe). Fig. 2(c) shows the current-induced hysteresis at $H_Z$ values of 0 Oe, +20 Oe, and +110 Oe, illustrating the change in the sign of the $V_H$ signal at +45 mA for $H_Z$ = +110 Oe compared to $H_Z$ = +20 Oe. In Fig. 2(d), the logic operations and their outputs are demonstrated: the output is 0 when both logic A and B are 0 ($I_{pulse}$ = +45 mA and $H_Z$ = +20 Oe); otherwise, the output is 1 for the remaining three scenarios (01, 10, and 11). Additionally, a total of 80 logic sequences were executed repeatedly to verify the consistency and reproducibility of the logic operations. Fig. 2(e) presents an alternative method for validating OR logic operations by selecting different input values for logic C and D. In this case, logic C is 0 at an $I_{pulse}$ value of +45 mA and 1 at -45 mA, while logic D is 0 at an $H_Z$ value of -20 Oe and 1 at +90 Oe. These conditions are detailed in the table in Fig. 2(e), with corresponding hysteresis curves shown in Fig. 2(f) for $H_Z$ field values of -20 Oe (bottom) and +90 Oe (top). The output for this additional OR operation is displayed in Fig. 2(g), resulting in a logic operation consistent with the one shown in Fig. 2(d) and confirming the OR logic truth table presented in Fig. 2(a).

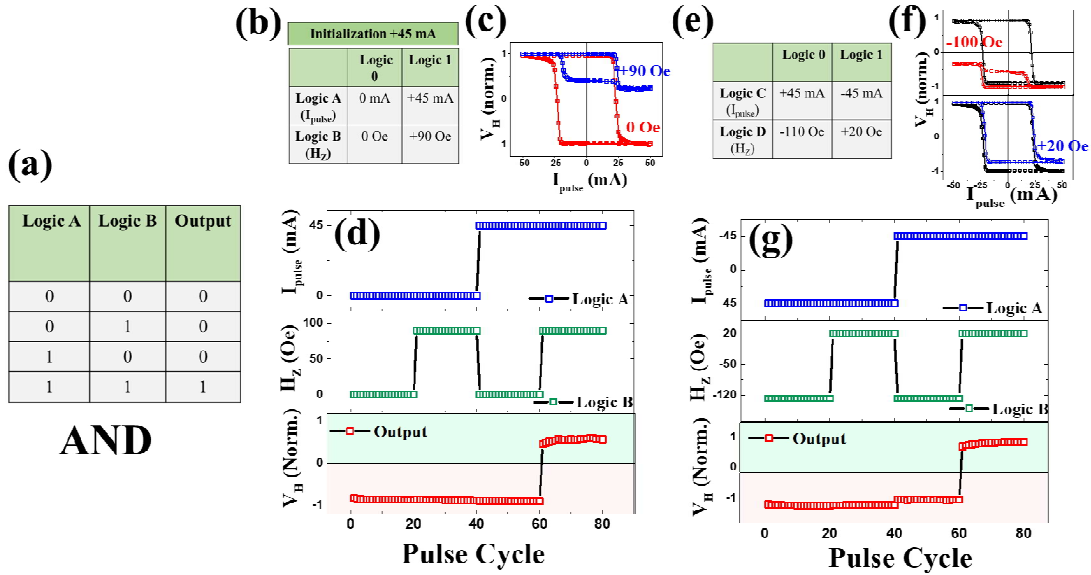

**Fig. 3(a)** AND gate truth table. **(b)** Input logic operations with an additional initialization step. **(c)** Current induced hysteresis at different field values mentioned in logic B **(d)** Operational sequence of AND gate with input condition mentioned in (b). **(e)** Input logic operations without additional initialization step. **(f)** Current induced hysteresis at different field values mentioned in logic D. **(g)** operational sequence of AND gate with input condition mentioned in (e).

Similar to the verification of OR logic operations, we validated AND logic operations (truth table in Fig. 3(a)) through the precise selection of input conditions. Fig. 3(b) illustrates one such set of input conditions, and Fig. 3(c) shows the current induced hysteresis under these $H_Z$ values. It is important to note that an initialization condition is required for this specific case. Before each operational sequence depicted in Fig. 3(d), an initial state with low Hall voltage is achieved by applying a +45 mA current pulse in the absence of $H_Z$. Additionally, the operational sequences were executed, and their outputs, shown in Fig. 3(d), follow the AND truth table. Moreover, selecting input logic values as demonstrated in Fig. 3(e) enables AND logic operations without the need for an initialization step. In this case, logic C and D are defined in Fig. 3(e), with the figures for SOT-induced hysteresis at various logic D inputs shown in Fig. 3(f). Fig. 3(g) demonstrates operational sequences spanning 80 cycles, adhering to the AND logic truth table depicted in Fig. 3(a). Furthermore, this method of $H_Z$-induced multistates can be employed to validate other logic operations, including NAND, NOR, and ALWAYS ON (refer to supplementary information). Thus, this approach not only enables reconfigurability across various logic operations but also facilitates reconfigurability within a specific spin logic operation in a single SOT-based memory device.

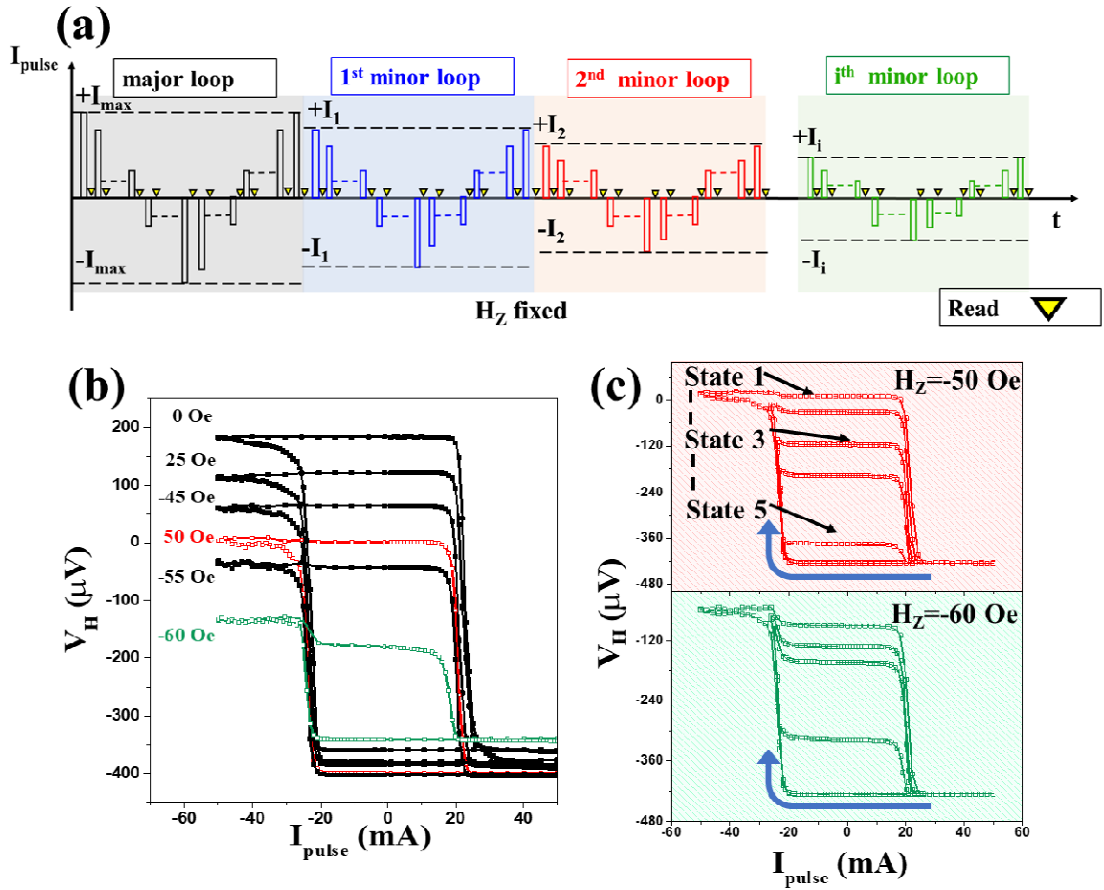

Fig. 4(a) Writing sequence for multiple minor loops at a fixed $H_Z$ field. Here, $I_{max}$=+60 mA and $I_1$,

$I_2...I_i$, are below critical current values.(b) Magnetic state in SOT induced switching controlled by $H_z$ field. (c) Applying the writing sequence at distinct hysteresis levels (specifically Hz = -50, and -60 Oe) and observing their multiple minor loop states.

In addition to performing logic operations, the availability of highly dense memory states within a single memory unit is crucial for in-memory computing applications. The density of these memory elements can be further increased n-fold by tracing minor loops of each current-induced hysteresis at different $H_z$ field values(Fig. 1). The writing procedure is illustrated in Fig. 4(a). To experimentally validate this process, Fig. 4(b) shows multiple current-induced hysteresis loops, demonstrating a reduction in magnetization state at -50 mA as the $H_z$ field magnitude increases. Among these various $H_z$ field-dependent hysteresis loops, we selected two major loops corresponding to $H_z$ field values of -50Oe (red) and -60 Oe (green) as a case study. At a fixed $H_z$ field, a current sweep is performed as the writing mechanism depicted in Fig. 4(a). The initial state is achieved by applying a $+I_{max}$ (+50 mA) current pulse, which has a higher magnitude than the critical switching current. Subsequently, a current pulse sweep is performed from $+I_i \rightarrow 0 \rightarrow -I_i \rightarrow 0 \rightarrow +I_i$. Here, the value of $I_i$ is less than the magnitude of $+I_{max}$ and varies in each sweep (i=1, 2, 3,...). These current sweeps with different values of $I_i$ produce the minor loops of the current-induced major hysteresis loops. In Fig. 4(c), minor loops associated with major hysteresis loops at $H_z$ field values of -50Oe (red) and -60 Oe (green) are presented, showing additional magnetic states depending on the choice of $I_i$. Each magnetic state in the minor loop corresponding to a particular $I_i$ of the multistate curve is attributed to the intermediate domain states. Notably, the density of these memory states for each hysteresis loop can be further enhanced by increasing the number of $I_i$ sweeps. Therefore, combining $H_z$ field-controlled current-induced switching with the minor loop method offers an n-fold enhancement in memory density within a single memory device.

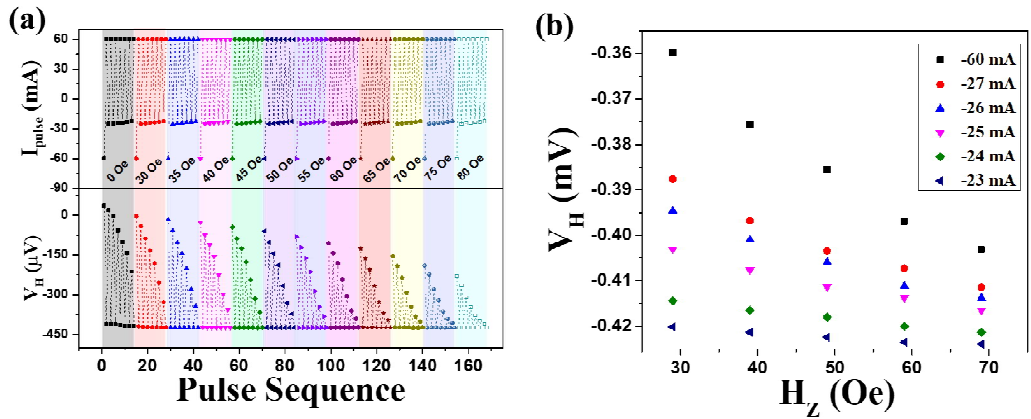

Fig. 5**(a)** Two-step writing sequence, in first step, a -60 mA current pulse fix one saturation state at fix $H_Z$; in second step, an initialization pulse of +60 mA is followed by applying current $I_i$ to stabilize the state dependent on the minor loop. **(b)** Hall voltage ($V_H$) as a function of $H_Z$ fields for multiple minor loop current ($I_i$) values.

Next, we combined $H_Z$-controlled SOT switching with $I_i$ controlled SOT minor loop methods to achieve high density of multistate memory. We propose a two-step writing mechanism to demonstrate this application. Initially, magnetic state corresponding to a fix $H_Z$ value is achieved using $H_Z$ integration with SOT current pulses, followed by employing various sweeping current pulses ($I_i$) to showcase numerous multi-states. This writing process, shown in the upper part of Fig. 5(a), starts with an initialization current pulse of +60 mA, which is essential for achieving the initial magnetic state with maximum -$V_H$ magnitude (refer to Fig. 1 and 4). In the bottom part of Fig. 5(a), this initial state corresponds to a Hall signal of ~ -425 μV. Following the initialization pulse, a -60 mA current pulse (larger than critical current value) is applied, which results in a magnetic state which depends on the $H_Z$ magnitude. Subsequently, varying magnitudes of current pulses ($I_i$) smaller than the critical switching current are applied (-25 mA, -24.5 mA, -24 mA, -23.5 mA, -23 mA, and -22.5 mA). Prior to these $I_i$ pulses, it should be noted that an initialization current pulse of +60 mA is applied to reestablish the initial magnetic state. The same writing procedures are carried out across multiple $H_Z$ values. As a result of this two-step writing process, we achieved 84 distinct magnetic states using 12 different $H_Z$ field values and 7 different $I_i$ current values (bottom part of Fig. 5(a)). Each $H_Z$ value stabilized a unique magnetic state, while varying $I_i$ current values further subdivided these states into stable magnetic states. Therefore, the total number of memory states (84 in Fig. 5(a)) can be easily enlarged by increasing the number of $H_Z$ and/or $I_i$ current values. Remarkably, we present the highest number of multi-states observed to date in a single SOT-based device, with potential to further increase this memory density. Fig. 5(b) shows the plot of Hall signal as a function of multiple $H_Z$ values for various current magnitude (different shape and color of data points). Here, for a constant $I_i$ current, SOT-induced magnetic states vary distinctly with different $H_Z$ and vice versa. This cycle-to-cycle variation of a state not only confirms the uniqueness of magnetic state in each $H_Z$ cycle but could also have applications in reconfigurable physically unclonable functions for data encryption applications[27-29].

In summary, we demonstrate reconfigurable logic operations in a single PMA-based SOT device through the integration of SOT and magnetic field. We successfully executed AND, OR, NOR, NAND, and Always ON logic operations in a single SOT device. Utilizing a two-step writing process, we achieved 84 distinct magnetic states by varying $H_Z$ fields (step 1) and $I_i$ current values (step 2), marking a significant advancement in SOT-based memory density. These results highlight the feasibility and potential of our approach for enhancing memory performance and enabling multifunctional logic operations within a single memory device for in-memory computing applications.